\DocumentMetadata{}
\documentclass[sigconf]{acmart}

\usepackage{bm}
\usepackage{amsfonts}
\usepackage{amsmath}
\usepackage{color}
\usepackage{graphicx}
\usepackage{subfigure}
\usepackage{booktabs}
\usepackage{multirow}
\usepackage{enumitem}
\usepackage{setspace}
\usepackage{colortbl}
\usepackage{pifont}
\usepackage{balance}
\usepackage{xspace}
\usepackage{array}
\usepackage{stfloats}
\usepackage{algpseudocode}
\usepackage[noend,ruled,linesnumbered]{algorithm2e}
\usepackage{makecell}
\usepackage{url}
\usepackage{CJKutf8}

\definecolor{myblue}{rgb}{0.2, 0.2, 0.9}

\begin{document}
\title{Survivors, Complainers, and Borderliners: Upward Bias in Online Discussions of Academic Conference Reviews} 

\author{Hangxiao Zhu}
\email{hangxiao@tamu.edu}
\orcid{0009-0007-1394-3410}
\affiliation{%
  \institution{Texas A\&M University}
  \city{College Station}
  \state{Texas}
  \country{USA}
}

\author{Yian Yin}
\email{yy994@cornell.edu}
\orcid{0000-0003-3018-4544}
\affiliation{%
  \institution{Cornell University}
  \city{Ithaca}
  \state{New York}
  \country{USA}
}

\author{Yu Zhang}
\email{yuzhang@tamu.edu}
\orcid{0000-0003-0540-6758}
\affiliation{%
  \institution{Texas A\&M University}
  \city{College Station}
  \state{Texas}
  \country{USA}
}

\begin{abstract}
Online discussion platforms, such as community Q\&A sites and forums, have become important hubs where academic conference authors share and seek information about the peer review process and outcomes. However, these discussions involve only a subset of all submissions, raising concerns about the representativeness of the self-reported review scores. In this paper, we conduct a systematic study comparing the review score distributions of self-reported submissions in online discussions (based on data collected from Zhihu and Reddit) with those of all submissions. We reveal a consistent upward bias: the score distribution of self-reported samples is shifted upward relative to the population score distribution, with this difference statistically significant in most cases. Our analysis identifies three distinct contributors to this bias: (1) \textit{survivors}, authors of accepted papers who are more likely to share good results than those of rejected papers who tend to conceal bad ones; (2) \textit{complainers}, authors of high-scoring rejected papers who are more likely to voice complaints about the peer review process or outcomes than those of low scores; and (3) \textit{borderliners}, authors with borderline scores who face greater uncertainty prior to decision announcements and are more likely to seek advice during the rebuttal period. These findings have important implications for how information seekers should interpret online discussions of academic conference reviews.
\end{abstract}



\begin{spacing}{0.96}
\maketitle

\section{Introduction}
Online discussion hubs, such as community Q\&A platforms (e.g., Quora \cite{wang2013wisdom} and Zhihu \cite{deng2020contributes}) and forums (e.g., Reddit \cite{yuan2025behavioral} and 4chan \cite{jokubauskaite2020generally}), have become important channels for people to share and obtain information. In recent years, the explosive growth of submissions to academic conferences (particularly in AI \cite{kim2025position}) has led to more online discussions about the peer review process of these conferences. In fact, for most conferences, when authors first learn their submission scores (e.g., at the start of the author rebuttal period), they do not have access to the score distribution of all submissions. Consequently, they may turn to online discussions to seek information that can inform their next steps, such as whether it is worth submitting a rebuttal and, if so, whether to adopt an aggressive or conservative rebuttal strategy. During this process, some authors share their own scores to solicit specific advice from other users, which in turn provides the community with additional reference examples.

Although there has been extensive research on information seeking and perception in online discussions \cite{upadhyay2019complexity,yang2019seekers,chang2020don,kumar2017army}, studies in the context of academic conference reviews remain scarce.
In particular, for any given conference, only a subset of all submissions is discussed online.
Authors who choose to self-report their review scores in online discussions may not be sufficiently representative to cover all score ranges proportionally, raising concerns about whether statistics derived from this subset are accurate or unbiased compared with the overall population.
Resolving this uncertainty is crucial, as it directly shapes how submission authors should interpret the information shared on these platforms.
Yet it remains unclear whether the bias exists, in which direction, and to what extent.
To give two general-domain examples that reveal biases in opposite directions, Dodds et al. \cite{dodds2015human} find that people tend to talk about pleasant items in online spaces, while Watson et al. \cite{watson2024negative} show that negative news articles are shared more to social media.

\vspace{1mm}
\noindent \textbf{Contributions.} To answer the question above, this paper presents, to the best of our knowledge, the first systematic study to compare the review score distribution of conference submissions self-reported by authors in online discussions (which we refer to as the \textit{samples}) with that of all submissions to the same conference (the \textit{population}). To be specific, we examine discussions of ICLR 2024, ACL 2024, EMNLP 2024, ICLR 2025, and ACL 2025 on Zhihu, a Chinese community Q\&A platform, and Reddit, a predominantly English-language forum. These conferences are all recent large AI venues that have triggered substantial online discussions, and for which the population review score distributions are publicly available. Our results reveal a prevalent \textit{upward bias}: In every case we examine, the review score distribution of submissions mentioned in online discussions is shifted upward relative to that of the population, with the difference being statistically significant in most cases (e.g., p-value $< 0.001$).

To understand the source of this upward bias, we further conducted a comprehensive set of experiments. Our analysis reveals a novel finding: the bias is not driven by a single group of users. Instead, at least three distinct types of users contribute to it. The first group comprises authors of accepted papers (i.e., ``\textit{survivors}''). In fact, if we estimate a conference’s acceptance rate based on self-reported acceptance/rejection results in online discussions, the derived rate is substantially higher than the population acceptance rate. This suggests that after decision notifications, authors of accepted papers are more inclined to share their good news, while authors of rejected papers are more likely to conceal the bad news. Since accepted papers generally have higher scores than rejected ones, the disproportionate discussion of accepted papers naturally skews the sample distribution upward relative to the overall population.

Second, among authors of rejected papers, those with high scores (i.e., ``\textit{complainers}'') are more likely to participate in discussions than those with low scores. Specifically, we compare the distribution of review scores for self-reported accepted submissions with that of all accepted submissions, and similarly compare self-reported rejected submissions with all rejected submissions. We find that the mean review score of accepted samples does not deviate much from that of the overall accepted population. By contrast, the mean review score of rejected samples ranks very high within the score distribution of all rejected submissions (the top 1\%-20\%). In other words, for accepted submissions, we find no evidence that authors of exceptionally high-scoring, award-worthy papers or those of borderline accepted papers are more inclined to join online discussions than the others. However, for rejected submissions, authors of high-scoring rejected papers are significantly more likely to participate in online discussions than those whose papers receive low scores. We also present two cases illustrating how authors of high-scoring rejected papers voice complaints about the review quality and process.

\begin{table*}[!t]
\caption{Statistics of collected data.}
\vspace{-1em}
\small
\begin{tabular}{ccccc}
\toprule
\textbf{Venue}               & \textbf{Available Population Information} & \textbf{Score(s)}               & \textbf{\begin{tabular}[c]{@{}c@{}}\# Submissions in\\ Zhihu Threads\end{tabular}} & \textbf{\begin{tabular}[c]{@{}c@{}}\# Submissions in\\ Reddit Threads\end{tabular}} \\
\midrule
ARR 2024 February (ACL 2024) & Score Distribution, Acceptance Rate & Review Score, Meta-Review Score & 212            & 99              \\
ARR 2024 June (EMNLP 2024)   & Score Distribution, Acceptance Rate & Review Score, Meta-Review Score & 106            & 117             \\
ARR 2025 February (ACL 2025) & Score Distribution, Acceptance Rate & Review Score, Meta-Review Score & 73             & 226             \\
ICLR 2024                    & Score Distribution, Acceptance Rate & Review Score                    & 147            & 110             \\
ICLR 2025                    & Score Distribution, Acceptance Rate & Review Score                    & 64             & 107             \\
\midrule
WWW 2025                     & Acceptance Rate                     & Novelty, Technical Quality      & \multicolumn{2}{c}{61 (Zhihu + Reddit)}           \\
KDD 2025                     & Acceptance Rate                     & Novelty, Technical Quality      & \multicolumn{2}{c}{48 (Zhihu + Reddit)}           \\
CVPR 2025                    & Acceptance Rate                     & Review Score                    & 174            & 115             \\
AAAI 2025                    & Acceptance Rate                     & Review Score                    & 90             & 136             \\
\bottomrule
\end{tabular}
\vspace{-0.5em}
\label{tab:dataset}
\end{table*}

Third, not all online discussion posts appear after decision notifications. For instance, some authors disclose their review scores at the start of the author rebuttal period to seek advice, even though they do not yet know the decision. Surprisingly, we find that even after excluding posts reporting acceptance/rejection outcomes (i.e., removing the survivor and complainer effects), an upward bias still persists in the remaining no-decision-reported samples. By examining the distribution of these samples within the population, we observe that they are disproportionately concentrated in the score range around the acceptance threshold (i.e., ``\textit{borderliners}'') compared with the population distribution. A possible explanation is that authors with borderline submissions face greater uncertainty than those with clearly high or low scores prior to decision notification, making them more inclined to turn to online forums for information and advice. Since the conferences we examine (as well as most good academic conferences) have an acceptance rate below 50\%, the borderline score range lies above the population median. Consequently, a higher concentration of samples from this range naturally produces an upward bias in the observed sample distributions.

In summary, this paper makes the following contributions:
(1) Motivated by the previous mixed findings in the general domain, we conduct a systematic study of potential biases in how authors of academic conference submissions self-report their review scores on online discussion platforms.
(2) Our quantitative and qualitative analyses provide robust, cross-venue, and cross-platform evidence of a clear upward bias.
(3) By leveraging data on user reports of paper acceptance, rejection, and cases with no decision disclosed, we further dissect the bias into three underlying mechanisms: survivors, complainers, and borderliners. 

\section{Data}
\label{sec:data}

\noindent \textbf{Selection of Conferences.} We first identify the academic conferences to be studied to enable comparison between sample and population score distributions. Intuitively, these conferences should meet two criteria. First, they need to be sufficiently large in scale, so as to trigger enough discussions on forums and community Q\&A platforms. This ensures that we can collect sufficiently many samples from online discussions to derive statistically significant results. At present, top-tier AI conferences (``AI'' in the broad sense) are more likely to meet this criterion. Second, population-level review score distribution for \textit{all} submissions must be publicly available. 
To the best of our knowledge, ICLR and ACL Rolling Review (abbreviated to ARR, which is used to unify the review process for top NLP conferences such as ACL, EMNLP, and NAACL) are the two venues that satisfy both criteria. In fact, ICLR publicly releases the review content, scores, and final decisions for all submissions, regardless of whether they were accepted, rejected, or voluntarily withdrawn by the authors. ARR, on the other hand, publishes the score distribution for all submissions on its official website\footnote{\url{https://stats.aclrollingreview.org/}}. Some other AI conferences, such as NeurIPS, typically only release the review scores for accepted papers and/or for rejected papers whose authors have opted to make the reviews public. As a result, we are unable to obtain complete and unbiased population-level data from these conferences.

Meanwhile, since some of our subsequent analyses do not require access to the review score distribution of a conference (e.g., in Section \ref{sec:survivors}, only the acceptance rate is needed, which is often available in conference proceedings), we broadened our scope beyond ICLR and ARR for such analyses. Specifically, we included WWW, KDD, CVPR, and AAAI. This expanded selection enables our study to encompass all major subfields of AI as defined by CSRankings\footnote{\url{https://csrankings.org/}}, thereby improving the generalizability of (some of) our findings.

\vspace{1mm}
\noindent \textbf{Getting Population Data.} For ARR, we select the three cycles with the highest submission volumes up to the time of our data collection (i.e., 2024 February, 2024 June, and 2025 February). These cycles correspond to the final submission rounds for ACL 2024, EMNLP 2024, and ACL 2025, respectively. We obtain the score distributions for these three cycles from ARR’s official GitHub repository\footnote{\url{https://github.com/acl-org/arr-health}}.
Two types of ``scores'' are available. The first is the average of the overall assessment (OA) scores given by the reviewers, ranging from 1 to 5 and rounded to the nearest 0.5 in the official ARR statistics (e.g., 3.17 is rounded to 3, and 3.33 to 3.5). For terminological convenience, we refer to this score as the \textit{average review score}. The second is the score given by the meta-reviewer, also ranging from 1 to 5. We refer to it as the \textit{meta-review score}. Since ARR is only responsible for reviewing and scoring submissions, while the final acceptance decision is made by the respective conferences, we use the acceptance rates calculated in the proceedings of ACL 2024, EMNLP 2024, and ACL 2025 as the population-level acceptance rates.

For ICLR, we consider the two most recent conferences: ICLR 2024 and ICLR 2025. We obtain the reviewer scores (which take values from the set $\{1, 3, 5, 6, 8, 10\}$) and acceptance status for each paper from the Paper Copilot \cite{yang2025paper} GitHub repository\footnote{\url{https://github.com/papercopilot/paperlists/tree/main/iclr}}. Using such information, we compute the average review score for each paper and derive the conference acceptance rates.

For WWW, KDD, AAAI, and CVPR, we consider the most recent edition of each conference (i.e., in 2025). As noted above, we are unable to obtain their population-level review score distributions. However, we retrieved the acceptance rates of these conferences from public sources (e.g., proceedings).

\vspace{1mm}
\noindent \textbf{Getting Sample Data from Online Discussions.} We consider two platforms commonly used by authors to discuss conference submissions and review outcomes: Zhihu and Reddit. Zhihu is a Chinese community Q\&A platform. For nearly every major AI conference, there is at least one discussion thread titled \begin{CJK}{UTF8}{gbsn}“如何看待 \{某会议\} 审稿意见？”\end{CJK} (i.e., ``\textsf{What do you think of the reviews for \{Conference\}?}''), \begin{CJK}{UTF8}{gbsn}“如何评价 \{某会议\} ？”\end{CJK} (i.e., ``\textsf{How would you evaluate \{Conference\}?}''), or similar\footnote{An example for ICLR 2025: \url{https://www.zhihu.com/question/660470115}}. Reddit, by contrast, is a predominantly English-language forum. Within the \texttt{r/MachineLearning} subreddit, discussion threads about academic conference reviews can be titled ``\textsf{\{Conference\} Paper Reviews Discussion}'', ``\textsf{\{Conference\} Paper Decisions}'', or similar\footnote{An example for ICLR 2025: \url{https://www.reddit.com/r/MachineLearning/comments/1gov5zd/d_iclr_2025_paper_reviews_discussion}}. On both Zhihu and Reddit, users can post answers and engage in follow-up conversations through nested replies. For the conferences mentioned earlier, we curate their corresponding discussion threads. Within each thread, we manually review all answers and nested replies to identify posts discussing the author's own review scores, meta-review scores, and/or paper decisions. One post may discuss multiple submissions, and Table \ref{tab:dataset} shows the number of paper submissions collected by us from Zhihu and Reddit threads for each conference.

A few additional points are worth noting here. First, not every post reports all aspects of the review data. For instance, some more recent posts only mention whether the paper is accepted or rejected without disclosing scores, while some earlier posts share scores but are not updated to include the final decision. Second, our data collection focuses on posts where users discuss the review scores or decisions of their own submissions. Posts reporting scores of others (e.g., papers they have reviewed or papers of their labmates) are excluded. Posts that merely promote an accepted paper without discussing its reviews are also not counted. Third, for WWW 2025 and KDD 2025, the numbers of relevant posts on Zhihu and Reddit are relatively small. To increase sample size for our analyses, we combine data from both platforms. Additionally, these two conferences do not use a single overall assessment score. Instead, reviewers provide separate scores for novelty and technical quality.

\section{Existence of the Upward Bias}
\label{sec:existence}

\begin{figure}[t]
\centering
\subfigure[Average Review Score, ACL Rolling Review]{
\includegraphics[width=0.98\linewidth]{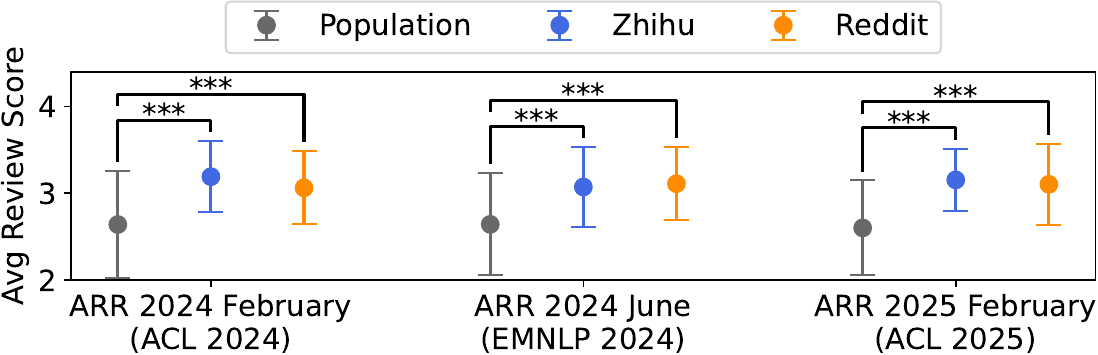}} \\
\vspace{-0.5em}
\subfigure[Meta-Review Score, ACL Rolling Review]{
\includegraphics[width=0.98\linewidth]{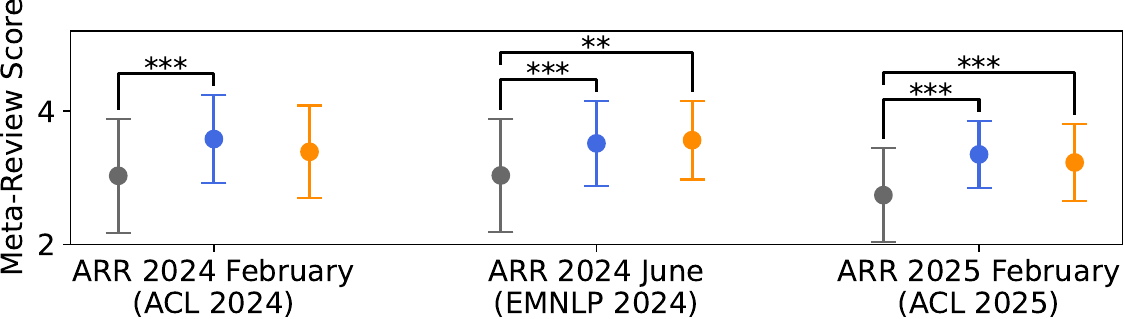}} \\
\vspace{-0.5em}
\subfigure[Average Review Score, ICLR]{
\includegraphics[width=0.98\linewidth]{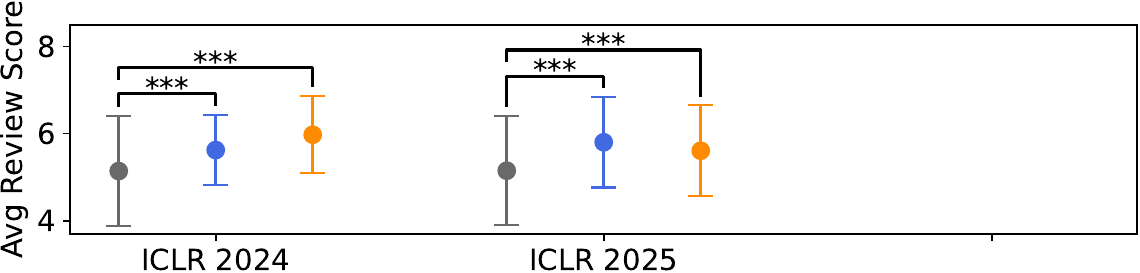}} \\
\vspace{-1em}
\caption{Comparisons between the population-level distribution of (meta-)review scores and the distribution reported by users on Zhihu/Reddit. *: p-value $<0.05$. **: p-value $<0.01$. ***: p-value $<0.001$. (*, **, and *** in the following figures have the same meanings as defined here.)}
\vspace{-1em}
\label{fig:all}
\end{figure}

The first question we aim to investigate is whether there is a significant difference between the population-level distribution of (meta-)review scores and the distribution reported by users on Zhihu/Reddit. Figure \ref{fig:all} compares population data with Zhihu/Reddit samples for ARR and ICLR.
For ARR, we examine the distributions of average review scores and meta-review scores. Since average review scores in population-level data are rounded to the nearest 0.5, we apply the same rounding to Zhihu and Reddit samples for consistency.
For ICLR, as meta-reviewers only provide a decision without numerical scores, we only explore the distribution of average review scores. In the error bar plots, solid dots represent the mean of each distribution, and error bars visualize the standard deviation.
Additionally, we conduct the Kolmogorov–Smirnov (KS) test \cite{smirnov1948table} to assess whether the sample distributions from Zhihu and Reddit significantly differ from the population distribution. The corresponding significance levels are marked in Figure \ref{fig:all}.

From Figure \ref{fig:all}, we observe a consistent upward bias in the review scores reported by users on Zhihu and Reddit, with their score distributions noticeably skewed higher compared to the corresponding population-level distribution. 
To be specific, in all cases, the mean of the sample distribution exceeds that of the population. For ARR's average review scores, the sample mean is, on average, 0.489 points higher than the population mean (i.e., a relative increase of 18.6\%). For ARR's meta-review scores, the sample mean is 0.503 points higher on average (i.e., a relative increase of 17.2\%). For ICLR’s average review scores, the sample mean is 0.608 points higher on average (i.e., a relative increase of 11.8\%). 
In nearly all cases, the upward bias of the sample distribution relative to the population is statistically significant, with p-values generally below 0.001. 
Moreover, the standard deviation of the sample distribution is always smaller than that of the corresponding population distribution.
These observations suggest that the Zhihu/Reddit samples are more likely drawn from one or more score subranges skewed above the population average. In other words, submissions within these higher-scoring subranges are more likely to be discussed by their authors on online forums.

\section{Sources of the Upward Bias}
Having recognized the presence of the upward bias, we now explore its underlying causes. Interestingly, we find that it is not driven by a single group of users. In fact, at least three distinct types of users contribute to this upward bias: ``survivors'', ``complainers'', and ``borderliners''.

\subsection{Survivors}
\label{sec:survivors}

Intuitively, one of the most straightforward explanations is the \textit{Survivorship Bias} \cite{mangel1984abraham,brown1992survivorship,elton1996survivor}. Authors of accepted papers (i.e., the ``survivors'') are more inclined to share their good news in online discussions. By contrast, authors of rejected papers may be less motivated to disclose disappointing outcomes. Instead, they may shift their focus to future conferences rather than revisiting threads related to the one that rejected them. Since accepted papers tend to have higher scores than rejected ones, a larger number of discussions around accepted papers naturally skews the Zhihu/Reddit sample distribution upward relative to the overall population.

\begin{figure}[t]
\centering
\subfigure[Main Conference Acceptance Rate, ACL Rolling Review]{
\includegraphics[width=0.98\linewidth]{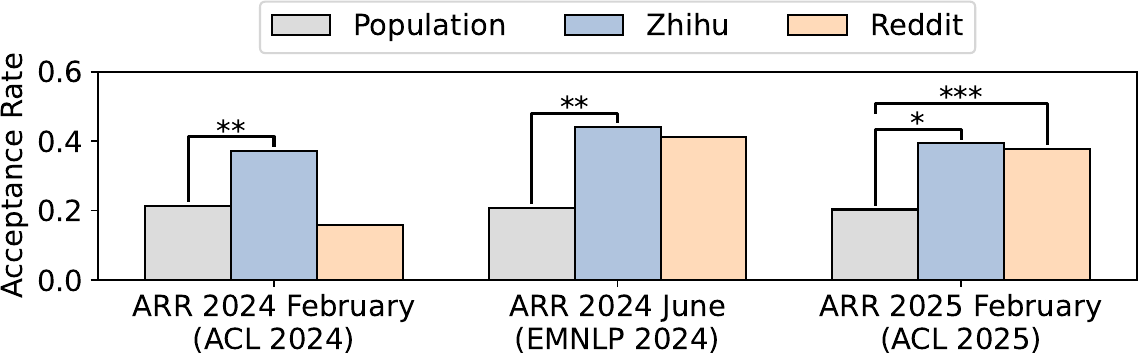}} \\
\vspace{-0.5em}
\subfigure[Main Conference + Findings Acceptance Rate, ACL Rolling Review]{
\includegraphics[width=0.98\linewidth]{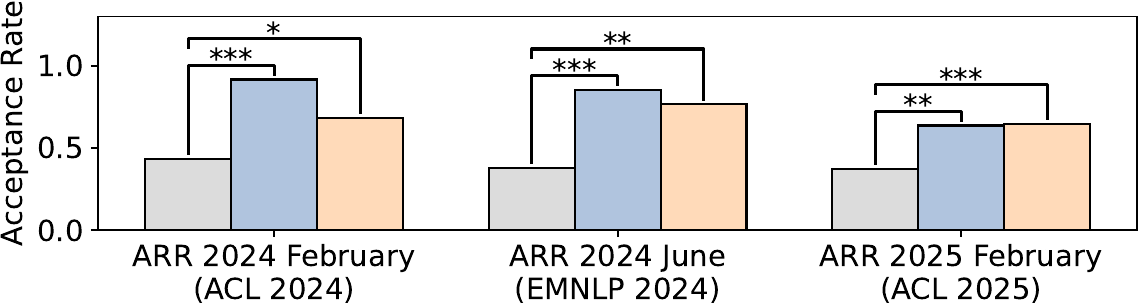}} \\
\vspace{-0.5em}
\subfigure[Acceptance Rate, ICLR]{
\includegraphics[width=0.98\linewidth]{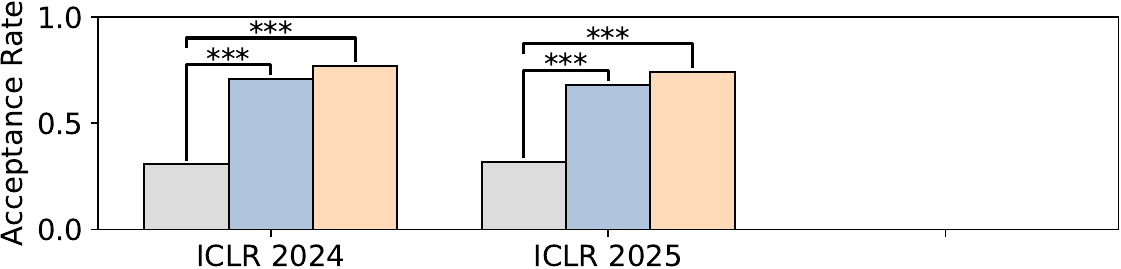}} \\
\vspace{-1em}
\caption{Comparisons between the population-level acceptance rate and the acceptance rate derived from Zhihu/Reddit posts for ARR and ICLR.}
\vspace{-0.5em}
\label{fig:acc}
\end{figure}

To validate the existence of the survivorship bias, we examine all Zhihu/Reddit posts in our collection that report the decision of the submission. For ARR, there are three possible outcomes: acceptance as a main conference paper, acceptance as a ``findings'' paper (which is considered slightly lower in quality by reviewers but still worthy of publication), and rejection. For ICLR and the other conferences listed in Table \ref{tab:dataset}, the decision is binary—either accepted or rejected. We also treat posts in which authors report voluntarily withdrawing their submission as equivalent to rejection.
Based on this, we compute the acceptance rate within the Zhihu/Reddit samples. For ARR, we report two acceptance rates: one for main conference papers and one for main conference + findings papers. For other conferences, we report a single acceptance rate. Figure \ref{fig:acc} compares the acceptance rates of the Zhihu/Reddit samples with the population-level acceptance rates for ARR and ICLR. We conduct a binomial test to examine whether the difference between the sample and population acceptance rates is statistically significant. Formally, the p-value indicates the likelihood that the observed accepted and rejected samples could have been drawn from a Bernoulli distribution with the population acceptance rate. The corresponding significance levels are marked in Figure \ref{fig:acc}.

As shown in Figure \ref{fig:acc}, with the only exception of the main conference acceptance rate for ARR 2024 February (where the acceptance rate derived from Reddit samples is lower than that of the population), all other cases show higher acceptance rates in the samples than in the population.
For Zhihu, the differences between the sample and population are statistically significant in all cases; for Reddit, in 6 out of 8 comparisons, the sample acceptance rate is significantly higher.
Across the 8 comparisons, the population's average acceptance rate is 30.4\%, while the average acceptance rates in the Zhihu and Reddit samples are 62.6\% and 56.9\%, respectively, which are 2.06 and 1.87 times the population's.

\begin{figure}[t]
\centering
\includegraphics[width=0.98\linewidth]{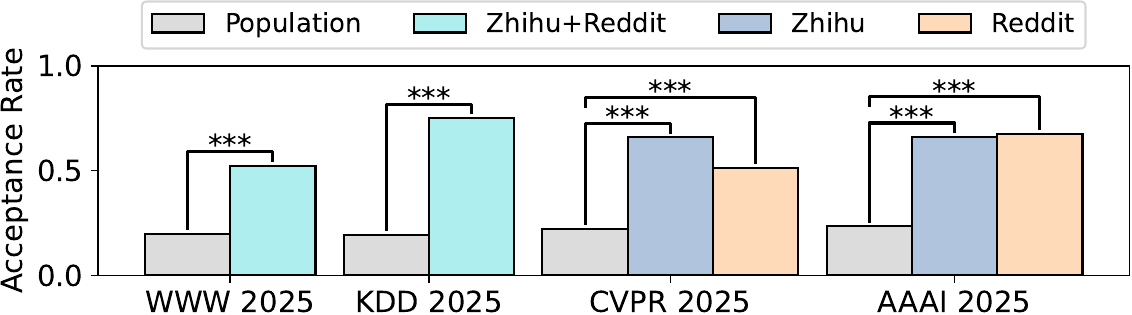}
\vspace{-0.5em}
\caption{Comparisons between the population-level acceptance rate and the acceptance rate derived from Zhihu/Reddit posts for WWW, KDD, CVPR, and AAAI.}
\vspace{-0.5em}
\label{fig:acc_other}
\end{figure}

\begin{table}[t]
\caption{Examples of posts from survivors. \textcolor{red}{WARNING: The content in this table may include potentially controversial or toxic views, which do not reflect the authors' stance.}}
\vspace{-1em}
\footnotesize
\begin{tabular}{p{8.1cm}}
\toprule
\cellcolor{black!10} \textbf{Example 1: ARR 2024 June (EMNLP 2024), Reddit} \\
\midrule
Got short main accept if the leaked one is correct. Meta:4 / OA: 4,4,4 / soundness: 4,4,4 / confidence: 5,4,4                        \\ \bottomrule \toprule
\cellcolor{black!10} \textbf{Example 2: ICLR 2025, Zhihu} \\
\midrule
\begin{CJK}{UTF8}{gbsn}
来还愿了！86665 -> 86666，accept.

\vspace{1mm}
第一次投 ICLR，主要是和 CVPR 太近，之前都是投 CVPR 为主。这次做热门方向，怕 CVPR 撞车太多，早一点下手。

\vspace{1mm}
个人感觉审稿水平明显高于 CVPR，审稿人都非常内行。有两个 contribution，一个很 solid 但是不太容易包装得很华丽，另一个比较水但是可以靠公式和可视化包装得很好看。结果五个审稿人都特别肯定那个不太好包装的 contribution，证明都是内行人。rebuttal 之后唯一负评也提分了，互动交流环节相比大多数顶会显得非常友好。
\end{CJK}

\vspace{1mm}
\textbf{(Translation)} Came back to share the good news! 86665 -> 86666, accept.

\vspace{1mm}
This was my first time submitting to ICLR. I mainly used to submit to CVPR, but since the deadlines were too close this time—and the topic is quite hot—I decided to try ICLR earlier to avoid too much competition at CVPR.

\vspace{1mm}
In my opinion, the review quality was noticeably higher than CVPR. All the reviewers were clearly knowledgeable. We had two main contributions: one was very solid but hard to make look flashy, and the other was relatively shallow but could be made to look impressive with some formulas and visualizations. Interestingly, all five reviewers strongly recognized the solid but hard-to-package contribution, which shows they really knew their stuff. After the rebuttal, even the only negative review raised their score. Overall, the discussion stage felt much friendlier and more constructive than most top-tier conferences.
\\ \bottomrule
\end{tabular}
\vspace{-0.5em}
\label{tab:eg_acc}
\end{table}

To demonstrate that survivorship bias is ubiquitous beyond ARR and ICLR, Figure \ref{fig:acc_other} presents a comparison between the population and sample acceptance rates for four additional conferences: WWW, KDD, CVPR, and AAAI.
The contrast between the sample and the population is more pronounced at these conferences. Every sample acceptance rate is significantly larger than the corresponding population acceptance rate with p-value $< 0.001$. On average, the population acceptance rate is 21.2\%, while the sample acceptance rate reaches 63.0\%—an astonishing 2.98 times the population rate.

To give a qualitative analysis, we present two examples of posts from survivors in Table \ref{tab:eg_acc}. One comes from Zhihu and the other from Reddit, covering ARR and ICLR respectively. Among them, the second post more clearly exemplifies survivorship bias through its celebratory tone. The author opens with ``\textit{Came back to share the good news}'' and then explains their strategic decision to submit to ICLR instead of CVPR, highlighting that it was a calculated move. Finally, the author praises the quality of ICLR reviews and even contrasts them favorably with those of CVPR, further reinforcing the narrative of a well-planned success.

\subsection{Complainers}
\label{sec:complainers}

\begin{figure}[t]
\centering
\subfigure[Average Review Score, ACL Rolling Review]{
\includegraphics[width=0.98\linewidth]{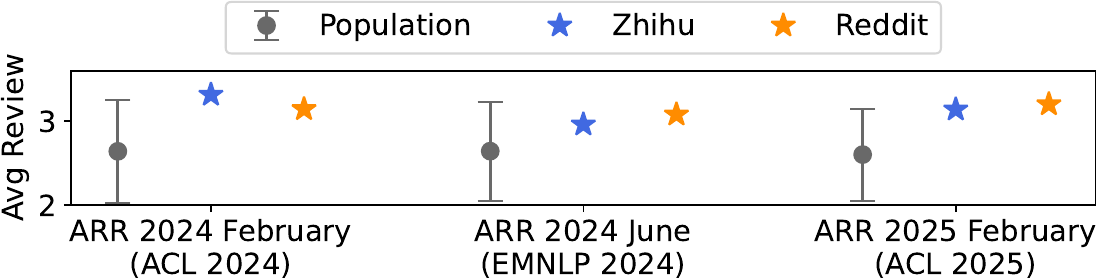}} \\
\vspace{-0.5em}
\subfigure[Meta-Review Score, ACL Rolling Review]{
\includegraphics[width=0.98\linewidth]{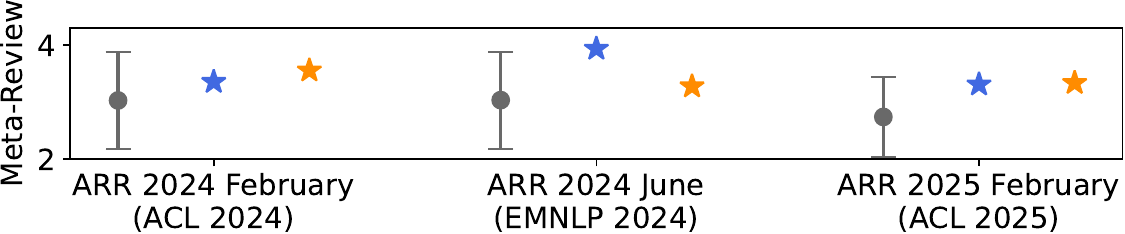}} \\
\vspace{-0.5em}
\subfigure[Average Review Score, ICLR]{
\includegraphics[width=0.98\linewidth]{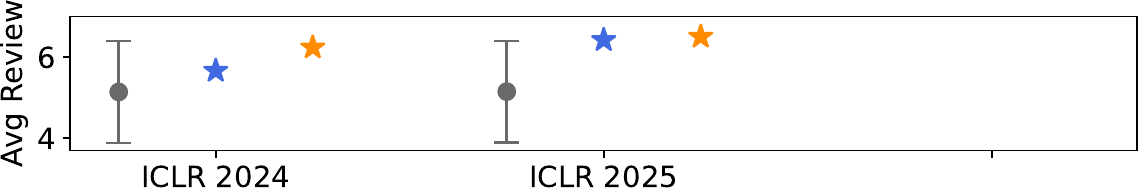}} \\
\vspace{-1em}
\caption{The weighted average score of samples reporting either acceptance or rejection, after being adjusted by the population acceptance rate. The adjusted sample mean still consistently exceeds the population mean.}
\vspace{-0.5em}
\label{fig:cali}
\end{figure}

\begin{figure}[t]
\centering
\subfigure[ICLR 2024, Accepted]{
\includegraphics[width=0.485\linewidth]{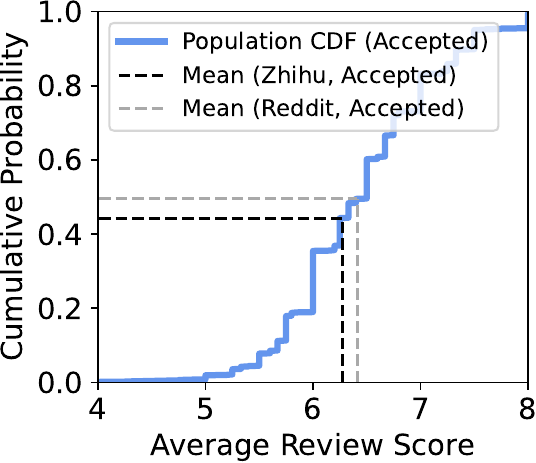}}
\subfigure[ICLR 2024, Rejected]{
\includegraphics[width=0.485\linewidth]{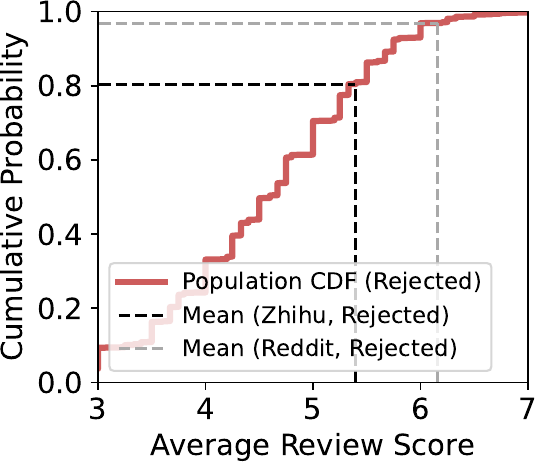}} \\
\subfigure[ICLR 2025, Accepted]{
\includegraphics[width=0.485\linewidth]{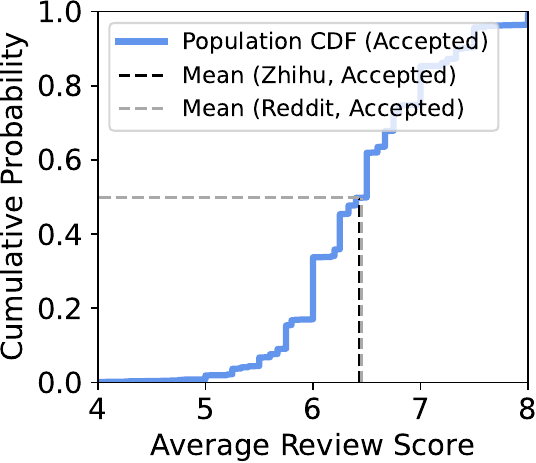}}
\subfigure[ICLR 2025, Rejected]{
\includegraphics[width=0.485\linewidth]{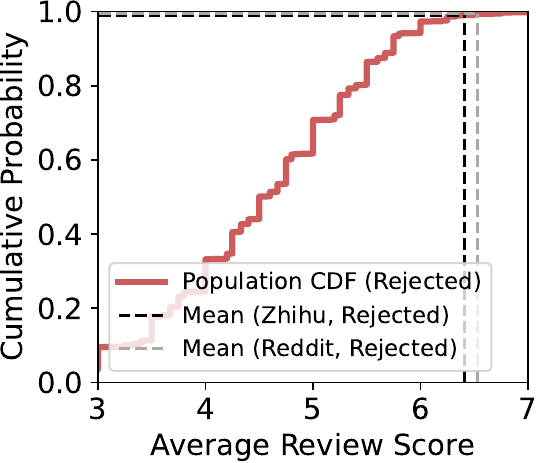}} \\
\vspace{-1em}
\caption{Population CDFs of the average review score for accepted/rejected papers at ICLR, with the mean values of the average review scores mentioned in Zhihu/Reddit posts where users report acceptance/rejection indicated by dashed lines.}
\vspace{-0.5em}
\label{fig:rej}
\end{figure}

The survivorship effect points to a clear compositional difference between online discussions and the population.
If this alone were sufficient to explain the upward bias, then a simple adjustment would be enough to completely eliminate the bias. We could compute a weighted average over the samples:
$\alpha\times \bar{s}_{\rm accept} + (1-\alpha)\times \bar{s}_{\rm reject}$, where $\bar{s}_{\rm accept}$ (resp., $\bar{s}_{\rm reject}$) is the average score of samples reporting acceptance (resp., rejection), and $\alpha$ is the \textbf{population} acceptance rate.
(For ARR, the weighted average is $\alpha\times \bar{s}_{\rm main} + \beta \times \bar{s}_{\rm findings} + (1-\alpha-\beta)\times \bar{s}_{\rm reject}$, where $\alpha$ and $\beta$ are population acceptance rates for main conference and findings papers, respectively.)
Yet interestingly, Figure \ref{fig:cali} shows that, even after such an adjustment, the scores reported online are still consistently higher than expected. This implies the existence of other effects and prompts us to ask: within the group of authors that got accepted (resp., rejected), who is more likely to \textit{share the good (resp., bad) news}?

To investigate this question, we consider the average review scores of accepted papers reported by Zhihu/Reddit users as well as those of rejected papers, examining how each set of scores is distributed within the corresponding population distribution. Figure \ref{fig:rej} presents the population cumulative distribution functions (CDFs) of the average review score for accepted and rejected papers at ICLR conferences\footnote{Here, we conduct the analysis for ICLR only. This is because ARR provides the score distribution across all submissions (both accepted and rejected) but does not disclose the acceptance status of individual papers. As a result, it is not possible to derive separate CDFs for accepted and rejected papers.}, shown in red and blue respectively. We compute the mean values of the average review scores mentioned in Zhihu and Reddit posts where users report acceptance/rejection, and mark these mean values in Figure \ref{fig:rej} with black and gray dashed lines. This highlights the quantile position of each mean value within the corresponding population distribution.

Surprisingly, there is a stark contrast in the quantile positions of average review scores reported by authors of accepted and rejected papers within their respective population distributions. For authors reporting acceptance, the mean value of the average review scores they share tend to lie around the 50\% quantile of all accepted papers. More precisely, across two ICLR conferences, the reported mean values from Zhihu and Reddit acceptance posts correspond to the 44.3\%, 49.5\%, 49.8\%, and 49.8\% quantiles of the population score distributions of accepted papers. In other words, for accepted papers, the mean value of the average review scores reported in online discussions is comparable to (if anything, slightly lower than) that of all accepted submissions. 
Hence the surviorship mechanism (Section \ref{sec:survivors}) appears largely uniform across the score distribution above the acceptance threshold:
authors of exceptionally high-scoring, award-worthy papers and those of borderline accepted papers are equally likely to participate in online discussions after decision notification.

\begin{table}[t]
\caption{Examples of posts from complainers. \textcolor{red}{WARNING: The content in this table may include potentially controversial or toxic views, which do not reflect the authors' stance.}}
\vspace{-1em}
\footnotesize
\begin{tabular}{p{8.1cm}}
\toprule
\cellcolor{black!10} \textbf{Example 1: ARR 2025 February (ACL 2025), Zhihu} \\
\midrule
\begin{CJK}{UTF8}{gbsn}
432，2 分给了 3 分的 soundness 和 3.5 的 excitement，argue 了一下他这个 2 分有问题，meta 最后给了 3 分 finding，结果出来 reject，实在是令人费解

\vspace{1mm}
邮件到了，死心了

\vspace{1mm}
问了一波，有 3 分被拒的，3.5 被拒的，4 分 finding 的，5 分 finding 的，还有2 分烂文靠学术圈的大手发力收了的

\vspace{1mm}
讲究一个纯随机

\vspace{1mm}
今年这届 acl 在我心里直接降级到 aaai 一个水平
\end{CJK}

\vspace{1mm}
\textbf{(Translation)} Got scores of 4, 3, and 2. The reviewer who gave a 2 also gave 3 for soundness and 3.5 for excitement. I argued that the 2 was unreasonable. The meta-reviewer ended up giving a 3 and recommending it for Findings. But the final decision was a reject—truly baffling.

\vspace{1mm}
The notification email came. I’ve given up hope.

\vspace{1mm}
I asked around: there are rejections with 3s and 3.5s, Findings acceptances with 4s and 5s, and even some garbage papers with 2s that got in thanks to big names in the academic community pushing for them.

\vspace{1mm}
It’s basically just pure randomness.

\vspace{1mm}
In my mind, this year’s ACL has officially dropped to the level of AAAI.
\\ \bottomrule \toprule
\cellcolor{black!10} \textbf{Example 2: ICLR 2024, Reddit} \\
\midrule
8666 - Reject... AC misread the important part...

\vspace{1mm}
is there any solution except for withdrawing the paper for another conference..
\\ \bottomrule
\end{tabular}
\vspace{-0.5em}
\label{tab:eg_rej}
\end{table}

In contrast, the situation is markedly different for rejected papers. The average review scores reported by authors of rejected papers on Zhihu and Reddit are evidently higher than the average scores of all rejected submissions. In fact, across ICLR 2024 and ICLR 2025, the reported mean values from Zhihu and Reddit rejection posts correspond to the 80.4\%, 96.8\%, 99.0\%, and 99.4\% quantiles of the rejected paper score distributions.
Notably, in Reddit posts related to ICLR 2025, the average review score of rejected papers reported by users is 6.527, which is even higher than the average score of accepted papers, 6.451, reported in the same forum! These findings suggest that authors of high-scoring rejected papers are significantly more likely to participate in online discussions than those whose papers received low scores.

The two examples in Table \ref{tab:eg_rej} offer a more concrete illustration of why authors of high-scoring rejected papers choose to post. Their primary goal is not to simply share bad news, but rather to express dissatisfaction with the decision and complain about their perceived quality of the reviews. For instance, the first Zhihu post about ARR uses phrases like ``\textit{truly baffling}'' and ``\textit{pure randomness}'', and even goes as far as criticizing other conferences (``\textit{this year’s ACL has officially dropped to the level of AAAI}''). The second Reddit post about ICLR bluntly states that ``\textit{the AC misread the important part}''. This tone of frustration and complaint stands in sharp contrast to the more celebratory tone observed in Table \ref{tab:eg_acc}.

\subsection{Borderliners}
\label{sec:borderliners}

\begin{figure}[t]
\centering
\subfigure[Average Review Score, ACL Rolling Review]{
\includegraphics[width=0.98\linewidth]{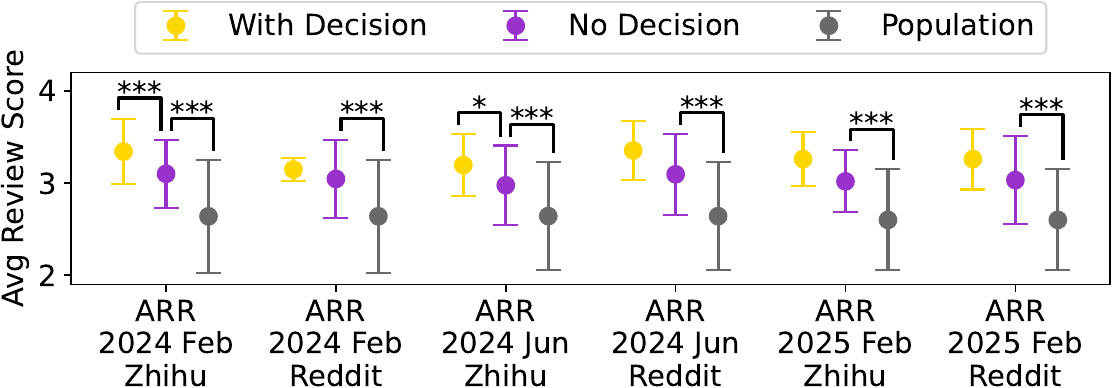}} \\
\vspace{-0.5em}
\subfigure[Meta-Review Score, ACL Rolling Review]{
\includegraphics[width=0.98\linewidth]{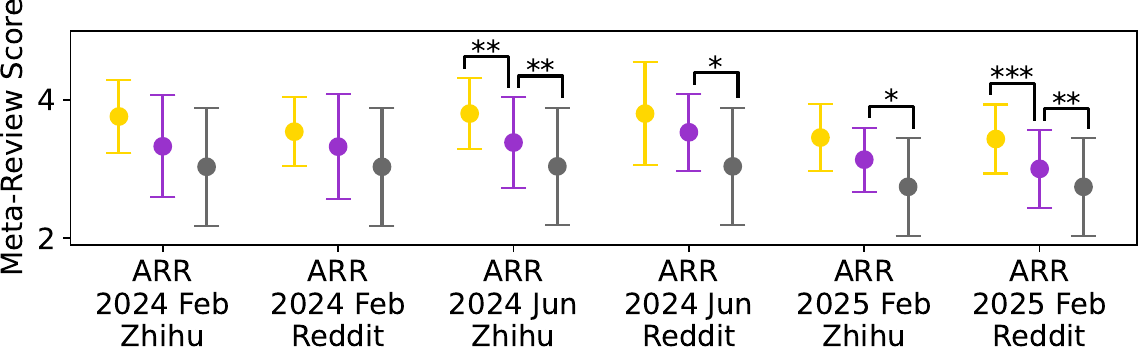}} \\
\vspace{-0.5em}
\subfigure[Average Review Score, ICLR]{
\includegraphics[width=0.98\linewidth]{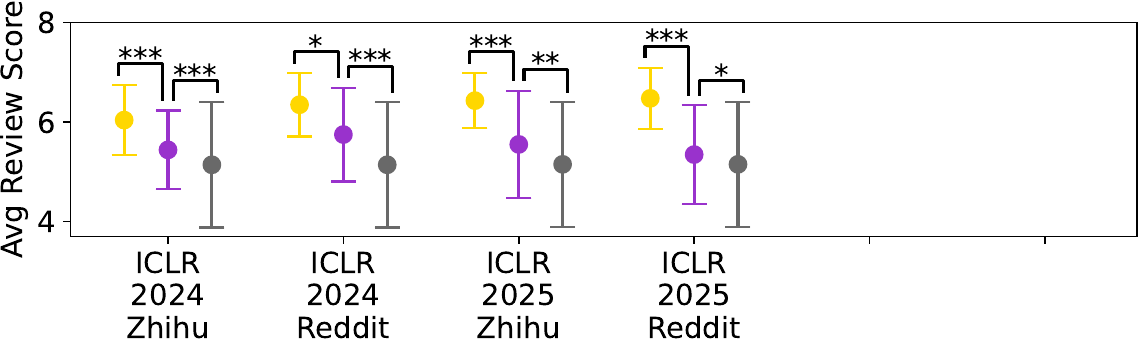}} \\
\vspace{-1em}
\caption{Comparisons among the score distribution from Zhihu/Reddit posts mentioning paper decisions, that from posts without paper decisions, and the population distribution for ARR and ICLR.}
\vspace{-0.5em}
\label{fig:res}
\end{figure}

In Sections \ref{sec:survivors} and \ref{sec:complainers}, we have discussed how posts reporting paper acceptance and rejection, respectively, contribute to the upward bias. A natural follow-up question is: How about posts only disclosing review scores without mentioning the final decision? For example, many online discussion threads of conference reviews emerge before decision notifications are released. At that point, authors do not yet know the outcome and can only discuss the review scores. In such cases, the previously mentioned survivor and complainer effects do not apply. So, do users who share scores before decisions are announced still contribute to the upward bias? And if so, what drives this behavior?

To explore this, we divide the Zhihu/Reddit posts into two categories: those reporting the paper decision and those not. For either category, we compute the mean and standard deviation of the reported (meta-)review scores and compare them to the corresponding population distributions. Figure \ref{fig:res} presents the comparisons. We again conduct the KS test to examine the statistical significance of differences between the decision-reported group and the no-decision-reported group, as well as between the no-decision-reported group and the overall population distribution.

\begin{figure*}[t]
\centering
\subfigure[Average Review Score, ACL Rolling Review]{
\includegraphics[width=0.98\linewidth]{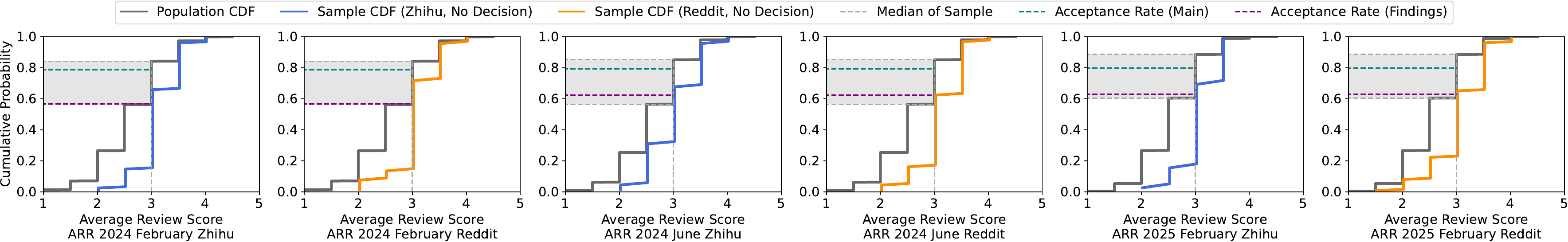}} \\
\subfigure[Average Review Score, ICLR]{
\includegraphics[width=0.655\linewidth]{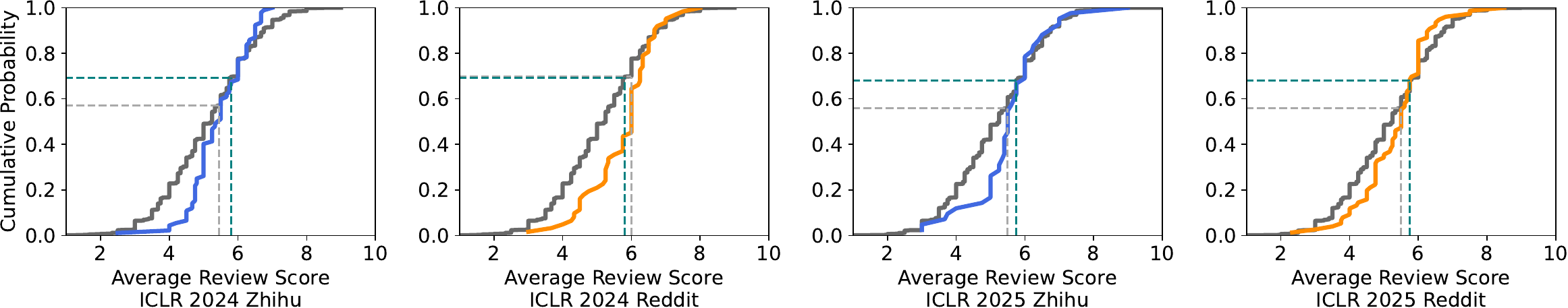}} \\
\vspace{-1em}
\caption{Comparisons between the population CDFs and the \textit{no-decision-reported} sample CDFs from Zhihu/Reddit for ARR and ICLR. We also highlight the median of each sample distribution and indicate the corresponding quantile that this median falls into within the population distribution.}
\vspace{-0.5em}
\label{fig:bord}
\end{figure*}

As shown in Figure \ref{fig:res}, across all cases, the no-decision-reported sample distribution is skewed higher than the population distribution: it has a higher mean and a smaller standard deviation. These distributional differences are statistically significant except for ARR 2024 February. This quantitatively demonstrates that even after removing survivorship and complainer effects, an upward bias still persists.
Specifically, the differences between the no-decision-reported sample mean and the population mean are 0.418 and 0.346 points for ARR’s average review and meta-review scores, and 0.376 points for ICLR’s average review scores.
However, when we compare these gaps with those calculated in Section \ref{sec:existence}, we find that the upward bias becomes smaller after removing the decision-reported group. In fact, as shown in Figure \ref{fig:res}, the mean scores of the no-decision-reported group are consistently lower than those of the decision-reported group. Moreover, in half of the cases, the differences between the decision-reported group and the no-decision-reported group are statistically significant.
In summary, there exists an effect within the no-decision-reported group that also introduces upward bias, but its magnitude appears smaller than that contributed by survivors and complainers. Let us now explore this effect in more detail.

Figure \ref{fig:bord} compares the population CDFs with the sample CDFs from Zhihu and Reddit, where we focus specifically on the posts that do not disclose acceptance/rejection (i.e., the no-decision-reported group). We also highlight the median of each sample distribution and indicate the corresponding quantile that this median falls into within the population distribution.
For ARR, since the average review scores are rounded to the nearest 0.5, the quantile corresponding to the median in the population is a range rather than a single point. In all ARR cases, the median score of the no-decision-reported group is 3.0. This score is not only the median but also the mode of the sample distribution (accounting for 48\% of the samples on average) and the result of rounding the sample mean to the nearest 0.5.
In the population distributions, a score of 3.0 corresponds to the top 15.8\%–43.8\% for ARR 2024 February, the top 14.9\%–43.6\% for ARR 2024 June, and the top 11.4\%–39.6\% for ARR 2025 February. In other words, the scores of the no-decision-reported group are concentrated within a range that spans from slightly above the main conference acceptance bar to around the main conference + findings acceptance rate. This range effectively represents submissions that are ``borderline'' in terms of the average review score.

For ICLR, we observe that the sample CDF lies below the population CDF in the low score range, but rises above it in the high score range. This indicates that the no-decision-reported sample is more concentrated around the middle score range compared to the population. The location of this concentration can be roughly characterized by the sample median, which corresponds to the top 43.0\%, 30.2\%, 44.2\%, and 44.2\% in the population distributions, respectively. In other words, the scores of the no-decision-reported group tend to cluster around or slightly below the acceptance bar, yet remain above the population median.

\begin{table}[t]
\caption{Examples of posts from borderliners.}
\vspace{-1em}
\footnotesize
\begin{tabular}{p{8.1cm}}
\toprule
\cellcolor{black!10} \textbf{Example 1: ARR 2024 February (ACL 2024), Zhihu} \\
\midrule
\begin{CJK}{UTF8}{gbsn}
evaluation track

\vspace{1mm}
3 3 3.5 不知道还有没有机会中主会，求保佑吧！

\vspace{1mm}
有没有同是这个 track 的小伙伴啊
\end{CJK}

\vspace{1mm}
\textbf{(Translation)} evaluation track

\vspace{1mm}
Got scores of 3, 3, and 3.5 — not sure if there’s still a chance for the main conference. Fingers crossed!

\vspace{1mm}
Anyone else in this track?
\\ \bottomrule \toprule
\cellcolor{black!10} \textbf{Example 2: WWW 2025, Reddit} \\
\midrule
I have 5/5/5/3/4 for technical and 5/4/3/4/4 for novelty. How are my chances? Anyone know any historical scores of papers that get in?
\\ \bottomrule
\end{tabular}
\vspace{-1em}
\label{tab:eg_bord}
\end{table}

Table \ref{tab:eg_bord} presents two representative examples from the no-decision-reported group. The first example, from ARR, reports an average review score of 3.17, which rounds to 3.0—the exact median and mode of the sample distribution shown in Figure \ref{fig:bord}. This is a borderline score, and the author expresses concern about the uncertainty with a ``\textit{Fingers crossed!}'' and attempts to connect with other users who submitted to the same track (since it is widely believed that different tracks vary in competitiveness and acceptance thresholds).
The second example, from WWW, reports average scores of 4.4 for technical quality and 4.0 for novelty (on a 1-7 scale), which is also a a borderline case. Therefore, the author asks about the likelihood of acceptance and seeks out others who received similar scores in previous years.

To summarize, during the period between receiving their review scores and the final decision notification, authors with borderline submissions often experience significant uncertainty. As a result, they turn to online discussion forums like Zhihu and Reddit as platforms for information seeking. By listening to others’ experience and advice, they hope to gain a more accurate estimate of their own paper’s chances of acceptance.
In contrast, authors with very high or very low scores face less uncertainty regarding the outcome and thus have less motivation to engage in such discussions. When a conference has an acceptance rate lower than 50\% (as is the case for most good CS conferences), the borderline score range will lie above the population median. Therefore, a higher concentration of samples from this range naturally leads to an upward bias in the observed sample distributions.

\section{Related Work}
\noindent \textbf{Analysis of Conference Peer Reviewing.}
With the explosive growth in submissions to CS (especially AI) conferences, recent studies have analyzed various aspects of the peer review process. For example, NeurIPS conducted two review consistency experiments in 2014 and 2021, where certain submissions were independently evaluated by two sets of reviewers \cite{cortes2021inconsistency,beygelzimer2023has}. The results reveal considerable randomness in review quality. In 2022, NeurIPS further asked authors and reviewers to rate the quality of peer reviews \cite{goldberg2025peer}. The results show that authors tend to rate reviews recommending acceptance of their papers more favorably, and that longer reviews are generally perceived as higher quality.
WSDM 2017 organizers carried out a single- versus double-blind review experiment \cite{tomkins2017reviewer}, confirming the presence of the Matthew effect \cite{merton1968matthew} in conference peer reviewing: papers from well-known authors and top institutions receive more favorable treatment under single-blind review. Liang et al. \cite{liang2024monitoring} observe an increasing use of ChatGPT in the peer review process of major AI conferences (e.g., ICLR, NeurIPS, CoRL, and EMNLP) since 2023. Su et al. \cite{su2025icml} use data collected from ICML 2023 to demonstrate the effectiveness of the isotonic mechanism \cite{su2021you}, where authors self-rank their own submissions (if there are multiple) to help de-noise reviewer-provided rankings.
Advanced NLP techniques have been explored for automatic paper-reviewer matching \cite{mysore2023editable,zhang2025chain}, a direction NeurIPS 2024 is also pursuing \cite{xu2024neurips}. The Paper Copilot platform \cite{yang2025paper} crowdsources conference review data (e.g., review scores) via Google Forms from authors to provide the community with more transparent statistics. Kim et al. \cite{kim2025position} propose several strategies (e.g., author feedback and reviewer rewards) in their position paper to address the ongoing peer review crisis in AI conferences.
These works offer valuable insights either by identifying issues in the current peer review system or by suggesting improvements. However, they primarily focus on the conferences themselves. By contrast, our study examines the interplay between conferences and online discussion forums and highlights an inherent issue in this linkage: the upward sample bias relative to the full population, which may result in misleading statistics for the broader community.

\vspace{1mm}
\noindent \textbf{Survivorship Bias and Positivity Bias.} A classic example of survivorship bias comes from military aviation \cite{mangel1984abraham}: when aircraft return from combat and some are lost, the surviving aircraft should be reinforced in areas that show no damage, as damage in those unseen areas is likely to have been fatal. This insight has since been extended to a variety of domains such as the predictability of future mutual fund performance based on past performance \cite{brown1992survivorship,elton1996survivor}, NIH grant applications for early-career researchers \cite{wang2019early}, and the selection of labeled data in information retrieval benchmarks \cite{gupta2022survivorship}.
Meanwhile, positivity bias \cite{boucher1969pollyanna} describes the tendency for people to talk about pleasant items. Previous studies \cite{dodds2015human} have shown its prevalence in online spaces.
However, to the best of our knowledge, survivorship and positivity biases in researchers’ reporting of academic paper reviews on social media remain unexplored. More importantly, this paper shows that the upward bias observed in online review discussions is not solely a result of the survivorship effect, but is also influenced by the behavior of complainers and borderliners.

\section{Conclusions and Discussions}
\label{sec:conclusion}
In this paper, we reveal the presence of an upward bias in how authors of academic conference submissions report their review scores on online discussion platforms such as Zhihu and Reddit. We observe that the distribution of scores observed from these user-reported samples is skewed higher compared to the actual population distribution. Through both quantitative and qualitative analyses, we identify three contributing factors to this bias: (1) authors of accepted papers (\textit{survivors}) are more likely to participate in discussions than those whose papers are rejected; (2) among rejected authors, those with high scores (\textit{complainers}) are more likely to engage than those with low scores; and (3) authors with scores near the acceptance threshold (\textit{borderliners}) are more likely to join discussions than others. We believe these findings can help guide submission authors in interpreting online review discussions more accurately.

We further discuss two additional considerations here. First, review scores for a submission may change dynamically during the rebuttal phase, and the scores captured in online discussions only reflect a specific point in time, rather than the final scores. However, it is generally understood that rebuttals, in a minority of cases, decrease the initial average score, so this factor is unlikely to explain the observed upward bias. Second, beyond the three types of users we identify, it remains unclear whether other users also contribute to the upward bias. Investigating whether similar patterns hold on other social media platforms or across different academic conferences is also a meaningful direction for future research. This requires access to open data and naturally motivates a call for conferences to release all reviews or at least review scores for public analysis.

\bibliography{arxiv_version}

\appendix
\section{Appendix}
\subsection{Data Sources}
\label{sec:source}

Table \ref{tab:source} lists the Zhihu and Reddit threads from which we collect online discussions of conference reviews, together with the sources of each conference’s acceptance rate. Notably, KDD 2025 has two submission cycles. For the analysis presented in Figure \ref{fig:acc_other}, we merge the samples from both cycles and compute the overall acceptance rate (as a weighted average based on the number of submissions in each cycle).

\begin{table*}[t]
\centering
\caption{Sources of collected data.}
\vspace{-1em}
\footnotesize
\begin{tabular}{lp{3.7cm}p{5.7cm}p{3.7cm}}
\toprule
\textbf{Venue} & \textbf{Zhihu Threads} & \textbf{Reddit Threads} & \textbf{Acceptance Rate Sources} \\
\midrule
ARR 2024 February (ACL 2024) & \url{https://www.zhihu.com/question/642309879} & \url{https://www.reddit.com/r/MachineLearning/comments/1boea3w/acl_2024_reviews_discussion/} & \url{https://aclanthology.org/2024.acl-long.0.pdf} \\
& & \url{https://www.reddit.com/r/MachineLearning/comments/1csqur3/d_acl_2024_decisions/} & \\
\midrule
ARR 2024 June (EMNLP 2024) & \url{https://www.zhihu.com/question/659192522} & \url{https://www.reddit.com/r/MachineLearning/comments/1ebmas6/d_acl_arr_june_emnlp_review_discussion/} & \url{https://aclanthology.org/2024.emnlp-main.0.pdf} \\
& & \url{https://www.reddit.com/r/MachineLearning/comments/1fkqxhh/d_emnlp_2024_results_notifications/} & \\
\midrule
ARR 2025 February (ACL 2025) & \url{https://www.zhihu.com/question/1888612458043790528} & \url{https://www.reddit.com/r/MachineLearning/comments/1jk6i69/d_acl_arr_feb_2025_discussion/} & \url{https://aclanthology.org/2025.acl-long.0.pdf} \\
& & \url{https://www.reddit.com/r/MachineLearning/comments/1kkynm9/d_acl_2025_decision/} & \\
\midrule
ICLR 2024 & \url{https://www.zhihu.com/question/622925909} & \url{https://www.reddit.com/r/MachineLearning/comments/17s9cnf/d_iclr_2024_paper_reviews/} & \url{https://github.com/papercopilot/paperlists/blob/main/iclr/iclr2024.json} \\
& & \url{https://www.reddit.com/r/MachineLearning/comments/196uyub/d_iclr_2024_decisions_are_coming_out_today/} & \\
\midrule
ICLR 2025 & \url{https://www.zhihu.com/question/660470115} & \url{https://www.reddit.com/r/MachineLearning/comments/1gov5zd/d_iclr_2025_paper_reviews_discussion/} & \url{https://github.com/papercopilot/paperlists/blob/main/iclr/iclr2025.json} \\
& & \url{https://www.reddit.com/r/MachineLearning/comments/1i5z6rd/d_iclr_2025_paper_decisions/} & \\
\midrule
WWW 2025 & \url{https://www.zhihu.com/question/666187176} & \url{https://www.reddit.com/r/MachineLearning/comments/1h56hno/d_www_2025_reviews_thewebconference/} & \url{https://dl.acm.org/doi/proceedings/10.1145/3696410} \\
\midrule
KDD 2025 & \url{https://www.zhihu.com/question/726358524} & \url{https://www.reddit.com/r/MachineLearning/comments/1fw7kga/d_kdd_2025_reviews/} & \url{https://dl.acm.org/doi/proceedings/10.1145/3690624} \\
& \url{https://www.zhihu.com/question/12035973262} & \url{https://www.reddit.com/r/MachineLearning/comments/1jrxh39/kdd_2025_cycle_2_reviews_are_out/} & \url{https://dl.acm.org/doi/proceedings/10.1145/3711896} \\
\midrule
CVPR 2025 & \url{https://www.zhihu.com/question/640949959} & \url{https://www.reddit.com/r/MachineLearning/comments/1i7dqlh/d_cvpr_2025_reviews/} & \url{https://cvpr.thecvf.com/Conferences/2025/News/Technical_Program} \\
& & \url{https://www.reddit.com/r/MachineLearning/comments/1ixpu28/d_cvpr_2025_final_decision/} & \\
\midrule
AAAI 2025 & \url{https://www.zhihu.com/question/657998175} & \url{https://www.reddit.com/r/MachineLearning/comments/1giqc9n/d_aaai_2025_phase_2_reviews/} & \url{https://aip.riken.jp/news/202412_aaai25} \\
& & \url{https://www.reddit.com/r/MachineLearning/comments/1h8kkjv/d_aaai_2025_phase_2_decision/} & \\
& & \url{https://www.reddit.com/r/MachineLearning/comments/1g1plva/d_aaai_2025_phase_1_decision_leak/} & \\
\bottomrule
\end{tabular}
\label{tab:source}
\end{table*}

\subsection{Additional Results on Decision-Reported Samples vs. No-Decision-Reported Samples}
\label{sec:resnores}

\begin{figure}[h]
\centering
\includegraphics[width=0.98\linewidth]{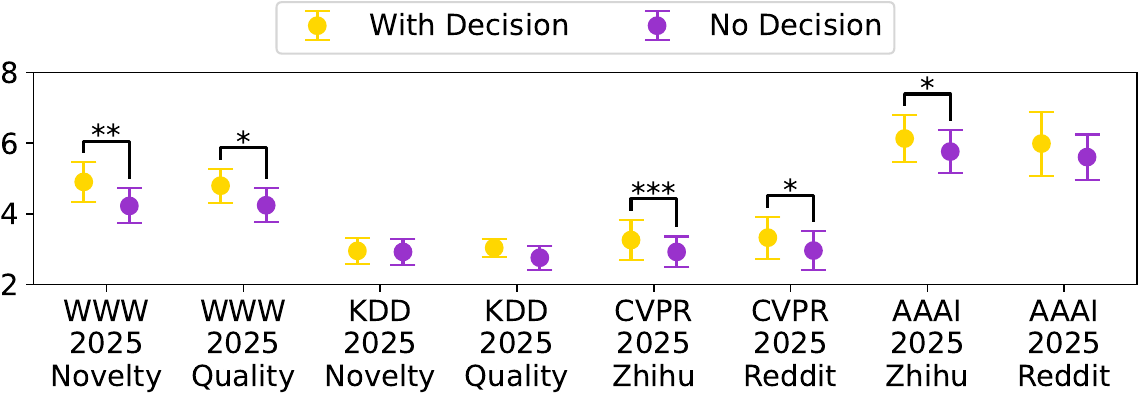}
\vspace{-0.5em}
\caption{Comparisons between the score distributions from Zhihu/Reddit posts with and without paper decisions for WWW, KDD, CVPR, and AAAI.}
\vspace{-0.5em}
\label{fig:res_other}
\end{figure}

In Section \ref{sec:borderliners}, we compare the decision-reported sample, the no-decision-reported sample, and the population score distributions for ARR and ICLR (Figure \ref{fig:res}). We find that the score distribution of the decision-reported sample is skewed higher than that of the no-decision-reported sample, which in turn is skewed higher than the population distribution. For other conferences in our collected data (i.e., WWW, KDD, CVPR, and AAAI), although we lack the population distribution, we can still compare the decision-reported and no-decision-reported samples, as shown in Figure \ref{fig:res_other}. Across all these comparisons, the decision-reported group consistently skews higher than the no-decision-reported group. In more than half of the cases, this difference is statistically significant, while in others (e.g., KDD) the gap becomes more subtle. One possible explanation for the latter is that KDD 2025 adopts a review score scale of \{1, 2, 3, 4\}, and we find that the vast majority of reported scores in online discussions are 2 or 3, which makes it difficult for the average scores of submissions to diverge meaningfully.

\subsection{Additional Results from Paper Copilot}
\label{sec:copilot}

\begin{figure}[h]
\centering
\includegraphics[width=0.98\linewidth]{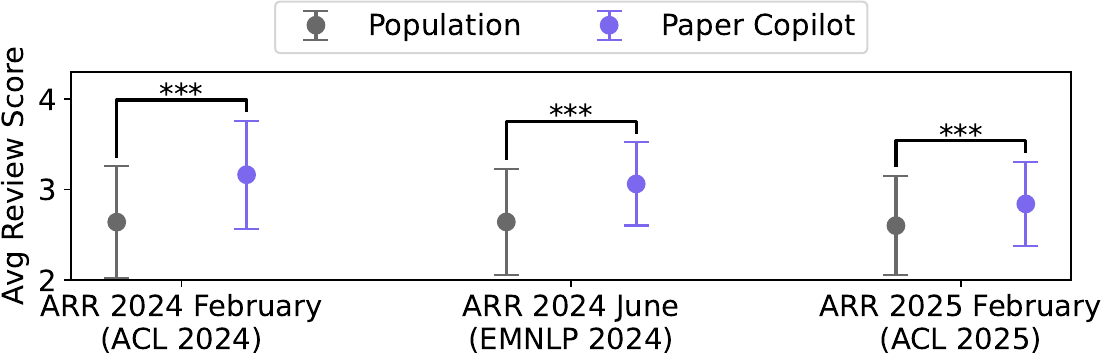}
\vspace{-0.5em}
\caption{Comparisons between the population-level distribution of average review scores and the distribution reported by authors via Paper Copilot \cite{yang2025paper}.}
\vspace{-0.5em}
\label{fig:copilot}
\end{figure}

In addition to self-reported review scores shared on online discussion platforms such as Zhihu and Reddit, the creator of the Paper Copilot website \cite{yang2025paper} also collects submission scores from conference authors via Google Forms and displayed them on the site, aiming to make the peer review process more transparent. This provides us with an additional relevant data source. Since Paper Copilot directly obtains the ICLR score distribution (i.e., the population distribution) from OpenReview rather than collecting data from authors, we only compare the self-reported sample distributions of ARR on Paper Copilot with the corresponding ARR population distributions. 
At the time we collected the data, Paper Copilot had gathered 211, 179, and 192 self-reported samples for ACL 2024, EMNLP 2024, and ACL 2025, respectively.
The comparisons are shown in Figure \ref{fig:copilot}. We observe that the upward bias in self-reported data remains consistent and significant. However, because Paper Copilot does not collect decision outcomes of submissions, we cannot determine whether the observed upward bias reflects the effects of survivors, complainers, or borderliners.

\subsection{Population Estimator with Biased Samples}
\label{sec:predictor}

\newcolumntype{C}[1]{>{\centering\arraybackslash}p{#1}}
\begin{table*}[t]
\caption{Earth mover’s distance between the Bias-Agnostic/Aware estimator and the true population distribution.}
\vspace{-1em}
\small
\begin{tabular}{cC{2cm}C{2cm}C{2cm}C{2cm}C{2cm}C{2cm}}
\toprule
& \multicolumn{2}{c}{\textbf{ARR 2024 Jun | ARR 2024 Feb}} & \multicolumn{2}{c}{\textbf{ARR 2025 Feb | ARR 2024 Jun}} & \multicolumn{2}{c}{\textbf{ICLR 2025 | ICLR 2024}} \\
& Zhihu                          & Reddit                        & Zhihu                          & Reddit                        & Zhihu                    & Reddit                  \\
\midrule
Bias-Agnostic Estimator & 0.1938                         & 0.1795                        & 0.1769                         & 0.1894                        & 0.1508                   & 0.1101                  \\
Bias-Aware Estimator    & \textbf{0.1000}                & \textbf{0.0515}               & \textbf{0.0038}                & \textbf{0.0453}               & \textbf{0.0449}          & \textbf{0.1061}        \\
\bottomrule
\end{tabular}
\label{tab:predictor}
\end{table*}

After revealing the presence of the upward bias, a natural follow-up question is whether this bias can be corrected to help researchers better understand their position within the conference submission landscape, thereby allowing authors to more accurately assess their competitive standing.
Let us consider the following scenario: a submission to a conference receives a review score $s$, and the authors wish to estimate the percentile of this submission within the overall population. Two approaches can be considered here:
\begin{itemize}[leftmargin=*]
\item \textbf{Bias-Agnostic}: The authors are unaware of the bias in online discussions and thus directly rely on the self-reported data from these discussions to derive the score distribution. We denote the CDF of this sample distribution as $f_{\rm sample}(\cdot)$. In this case, the estimated percentile is $f_{\rm sample}(s)$.
\item \textbf{Bias-Aware}: The authors are aware of the bias in online discussions. However, since the population CDF for the current conference $f_{\rm population}(\cdot)$ has not yet been released, they instead refer to the sample CDF $g_{\rm sample}(\cdot)$ and population CDF $g_{\rm population}(\cdot)$ from the previous iteration of the same conference (assuming that different iterations adopt the same scoring scale), and use mapping and inverse mapping to eliminate the bias. Formally, the estimated percentile is $g_{\rm population}(g_{\rm sample}^{-1}(f_{\rm sample}(s)))$.
\end{itemize}
Now, we need to compare the errors of these two estimators against the true population distribution $f_{\rm population}(\cdot)$, which can be measured using the earth mover’s distance (EMD) \cite{rubner1998metric}. Based on the data we collected, we consider the following three settings:
\begin{itemize}[leftmargin=*]
\item \textbf{ARR 2024 Jun | ARR 2024 Feb}: Predicting the population distribution of ARR 2024 June, where ARR 2024 February is considered as the previous iteration.
\item \textbf{ARR 2025 Feb | ARR 2024 Jun}: Predicting the population distribution of ARR 2025 February, where ARR 2024 June is considered as the previous iteration.
\item \textbf{ICLR 2025 | ICLR 2024}: Predicting the population distribution of ICLR 2025, where ICLR 2024 is considered as the previous iteration.
\end{itemize}
Table \ref{tab:predictor} demonstrates the results, from which we can observe that the Bias-Aware estimator consistently outperforms the Bias-Agnostic estimator. This suggests that while online discussions of academic conference reviews exhibit an upward bias, as long as we are aware of this bias, we can apply simple methods to partially correct it. Meanwhile, it is important to note that the Bias-Aware estimator still relies on the population score distribution from the previous iteration of the conference. This echoes our call in Section \ref{sec:conclusion} for releasing all review scores to promote a more transparent peer-reviewing process when wrapping up each iteration.

\end{spacing}
\end{document}